\documentclass[
aps,pra,reprint,
amssymb, amsfonts, amsmath,
eqsecnum,
nofootinbib
]{revtex4-2}
\usepackage[utf8]{inputenc}
\usepackage[T1]{fontenc}
\usepackage{dsfont,bm,wasysym,amsbsy,mathtools}
\usepackage{graphicx}
\usepackage{hyperref}
\usepackage[capitalize]{cleveref}
\crefname{section}{Sec.}{Secs.}
\usepackage{tcolorbox}
\usepackage{qcircuit,physics}
\usepackage[english]{babel}
\usepackage[hats]{./quantum_hqcs}
\usepackage{soul}
\hypersetup{
    colorlinks=true,
    linkcolor=blue,
    urlcolor=blue,
    citecolor=blue,
}

\newcommand*\Bell{\ensuremath{\boldsymbol\ell}}
\newcommand{\q}{\ope{q}}

\newcommand{\rotmat}{\op{R}_{L}^{z}}
\newcommand{\rotop}{\op{R}^{z}}
\renewcommand{\vec}[1]{\bm{\mathrm{#1}}}

\newcommand{ \CZL }{\op{\text{C}}{}_L^Z}

\newcommand{ \CZone }{ \CZ[ \tfrac{\pi}{\alpha^2} ] }
\newcommand{ \CZint }{ \op{\mathrm{C}}_\text{int} }
\newcommand{ \CZg }{ \op{\mathrm{C}}_G }

\renewcommand{\l}{\op{\ell}}
\newcommand{\mGop}[1][]{\op{m}_{G #1}}
\newcommand{\uGop}[1][]{\op{u}_{G #1}}

\newcommand{\GKP}{\mathrm{GKP}}
\newcommand{\Hilbert}{\mc{H}}
\newcommand{\CV}{\text{CV}}

\newcommand{\intalpha}{
\int_{-\alpha/2}^{+\alpha/2}\hspace{-4pt}
}

\begin{document}


\usetikzlibrary{shapes,fit,positioning}
\usetikzlibrary{quotes}

\definecolor{airforceblue}{rgb}{0.36, 0.54, 0.66}
\definecolor{amaranth}{rgb}{0.73, 0.26, 0.26}
\definecolor{azure}{rgb}{0.39, 0.55, 0.88}


\tikzstyle{mgkp}=[
    circle,draw,thick,dashed,amaranth,minimum width=20
]

\tikzstyle{mcircle}=[
    circle,draw,thick,dashed,amaranth,minimum width=13,minimum height=13
]

\tikzstyle{mrectangle}=[
    rectangle,thick,draw,dashed,amaranth,minimum width=12,minimum height=12
]

\tikzstyle{emptyrectanglenode}=[
    draw,
    rectangle,
    minimum width=12,
    minimum height=12
]
\tikzstyle{emptycirclenode}=[
    draw=black,
    fill=none,
    circle,
    minimum width=13
]
\tikzstyle{rectanglenode}=[
    draw=black,
    fill=amaranth,
    rectangle,
    minimum width=12,
    minimum height=12
]
\tikzstyle{circlenode}=[
    draw=black,
    fill=amaranth,
    circle,
    minimum width=13
]
\tikzstyle{cvnode}=[
    circle,
    draw=black,
    fill=black,
    minimum width=20
]
\tikzstyle{emptycvnode}=[
    circle,
    draw,
    minimum width=20,
    line width=1
]
\tikzstyle{logicalnode}=[
    draw=black,
    fill=azure,
    diamond,
    minimum width=15,
    minimum height=15
]
\tikzstyle{emptylogicalnode}=[
    draw=black,
    diamond,
    minimum width=15,
    minimum height=15
]
\tikzstyle{cvloop}=[
    loop,
    min distance=3em,
    looseness=6,
    line width=0.7,
]
\tikzstyle{rgaugeloop}=[
    line width=0.7,
    loop,
    in=-45, out=45,
    min distance=3em,
    looseness=6
]
\tikzstyle{lgaugeloop}=[
    line width=0.7,
    in=-135, out=135,
    loop,
    min distance=3em,
    looseness=6
]
\tikzstyle{logicalloop}=[
    line width=0.7,
    loop,
    min distance=3em,
    looseness=6
]
\tikzstyle{cvline}=[
    line width=1.2,
]
\tikzstyle{weight}=[
    fill=white,
    draw,
]
\usetikzlibrary{shapes,fit}
\usetikzlibrary{positioning}
\tikzstyle{gaugenode}=[draw, fill, color=black, circle, minimum width=8]
\tikzstyle{gkpgaugenode}=[draw, color=black, circle, minimum width=8]
\tikzstyle{gaugenodemeas}=[draw, black, circle, dashed, minimum width=8]
\tikzstyle{gaugerectanglenode}=[draw, black, loosely dotted, rectangle]
\tikzstyle{logicalrectanglenode}=[draw, black, loosely dotted, rectangle]
\tikzstyle{gaugeline}=[black]
\tikzstyle{logicalline}=[black]
\tikzstyle{redline}=[amaranth]
\tikzstyle{doubleline}=[double,black]
\tikzstyle{logicalcz}=[black, very thick]

\tikzstyle{dvnode}=[
    circle,
    draw=black,
    minimum width=15
]

\title{Hidden qubit cluster states}

\author{Giacomo Pantaleoni}
\email{gpantaleoni@null.net}
\affiliation{Centre for Quantum Computation \& Communication Technology, School of Science, RMIT University, Melbourne, VIC 3000, Australia}
\author{Ben Q. Baragiola}
\affiliation{Centre for Quantum Computation \& Communication Technology, School of Science, RMIT University, Melbourne, VIC 3000, Australia}
\author{Nicolas C. Menicucci}
\affiliation{Centre for Quantum Computation \& Communication Technology, School of Science, RMIT University, Melbourne, VIC 3000, Australia}

\begin{abstract}
Continuous-variable cluster states (CVCSs) can be supplemented with Gottesman-Kitaev-Preskill (GKP) states to form a hybrid cluster state with the power to execute universal, fault-tolerant quantum computing in a measurement-based fashion.  As the resource states that comprise a hybrid cluster state are of a very different nature, a natural question arises:
Why do GKP states interface so well with CVCSs?  To answer this question, we apply the recently introduced subsystem decomposition of a bosonic mode, which divides a mode into logical and gauge-mode subsystems, to three types of cluster state: CVCSs, GKP cluster states, and hybrid continuous-variable (CV)-GKP cluster states.  We find that each of these contains a ``hidden'' qubit cluster state across their logical subsystems, which lies at the heart of their utility for measurement-based quantum computing.  To complement the analytical approach, we introduce a simple graphical description of these CV-mode cluster states that depicts precisely how the hidden qubit cluster states are entangled with the gauge modes, and we outline how these results would extend to the case of finitely squeezed states. This work provides important insight that is both conceptually satisfying and helps to address important practical issues like when a simpler resource (such as a Gaussian state) can stand in for a more complex one (like a GKP state), leading to more efficient use of the resources available for CV quantum computing.
\end{abstract}

\date{\today}

\maketitle

\section{ Introduction }
Measurement-based models of quantum computation~\cite{knill2001schemeefficient,raussendorf_one-way_2001,raussendorf_computational_2001,briegel_persistent_2001,nielsen_fault-tolerant_2005,nielsen_cluster-state_2006,menicucci_universal_2006} are appealing avenues towards fault-tolerant, universal quantum computing as they do not require active control over interactions from nonlinear effects.  In the context of cluster-state quantum computing~\cite{nielsen_optical_2004,nielsen_fault-tolerant_2005,nielsen_cluster-state_2006, menicucci_universal_2006}, gates are implemented by performing local, adaptive measurements on an entangled state that is prepared in advance.  This procedure is simple in continuous-variable cluster-state (CVCS) quantum computing~\cite{menicucci_universal_2006,gu_more_2009}, where homodyne measurements on a CVCS, which is built from squeezed momentum states~\cite{larsen_deterministic_2019, asavanant_generation_2019}, are used to implement any Gaussian unitary~\cite{van_loock_building_2007,alexander_measurement-based_2017,asavanant2020onehundred, larsen2020deterministic}.
Extending to universal continuous-variable (CV) operations requires at least one non-Gaussian resource, and even then the scheme is not necessarily compatible with error correction as the latter requires discretized quantum information.  One solution to both these problems is to supplement the CVCS with a bosonic code, which encodes digital quantum information into the CV mode~\cite{chuang_bosonic_1997,cochrane_macroscopically_1999,guillaud_repetition_2019,michael_new_2016,grimsmo_quantum_2020}. This has the additional benefit that universal, \emph{discrete}-variable quantum computation can be straightforwardly
 performed---\emph{i.e.},~the suite of quantum algorithms designed for qubits can be implemented within the the continuous-variable Hilbert space of a bosonic mode.

Promising proposals for measurement-based, fault-tolerant, universal quantum computation with CV modes~\cite{menicucci_fault-tolerant_2014,larsen_fault-tolerant_2021,bourassa2021blueprint} take advantage of a specific bosonic code---the Gottesman-Kitaev-Preskill (GKP) code, known for its resilience to common forms of CV noise, including displacement noise~\cite{gottesman_encoding_2001} and bosonic excitation loss~\cite{albert_performance_2018}.
These proposals rely on the interplay between CVCSs and the GKP encoding, each using as the entangled computational resource state a \emph{hybrid cluster state}, where some of the nodes in a CVCS are replaced with specific GKP states.
Thus, in this setting GKP states play two roles:
(1)~they serve as the carriers of qubit-encoded quantum information, and (2)~
they enable CV error correction~\cite{menicucci_fault-tolerant_2014, walshe_2020}, which is the foundation for fault tolerance.

However, it is still puzzling that CVCSs---Gaussian CV states with no explicit qubit encoding---act as the substrate for discrete-variable quantum computing.
Why is it the GKP code and not some other bosonic code that interfaces so well with CVCSs to the point where CVCS nodes may be freely replaced with GKP states? Conversely, why is it that in proposals based on GKP cluster states~\cite{bourassa2021blueprint,pantaleoni_modular_2020}, some of the modes may be replaced with squeezed states? A major factor is that the GKP logical controlled-$Z$ gate is the entangling operation used to stitch together momentum squeezed into a canonical CVCS~\cite{menicucci_universal_2006,gu_quantum_2009} and GKP states into a GKP cluster state
~\cite{fukui_high-threshold_2019,fukui_high-threshold_2018, yamasaki2020polylogoverhead}. Also, the GKP code is the only known code for which the Clifford gate set is realized via Gaussian unitaries~\cite{gottesman_encoding_2001}, which homodyne detection implements on a CVCS. Moreover, this can be extended to fault-tolerant universality without additional non-Gaussian resources~\cite{baragiola_all-gaussian_2019,yamasaki2020cost}. A new perspective is given by a \emph{subsystem decomposition (SSD)} of a bosonic mode based on \emph{modular position}~\cite{pantaleoni_modular_2020}. In the SSD picture, which divides a single CV mode's Hilbert space into that of a logical qubit and a remaining gauge mode, \emph{any} CV state contains a logical-qubit state. Each infinitely squeezed momentum node of an ideal CVCS contains a logical-qubit $\ket{+}$ state, the same logical-subsystem state that an ideal $\ket{+_\text{GKP}}$ does~\cite{pantaleoni2021}.

We take these single-mode state decompositions and an SSD of the entangling CV controlled-$Z$ gate to inspect the logical content of CVCSs and of the larger class of hybrid cluster states, including the limiting case of a GKP cluster state.  We find that all hybrid cluster states contain a ``hidden'' qubit cluster state in their logical subsystems, whose entanglement with the gauge subsystems is restricted to the neighborhood of the non-GKP nodes in the cluster. We introduce a simple graphical description for hybrid cluster states that reveals this entanglement structure, and with it, we show how a logical GKP qubit teleports successfully along a linear CVCS via successive measurements, with the particular form of the GKP state serving to ``unzip'' the logical cluster state from the gauge modes at each measurement step.

This paper is structured as follows. In \cref{sec:ssd}, we briefly review the subsystem decomposition we intend to use, and we apply it to the various constituents of ideal hybrid cluster states---the single-mode states at each node (infinitely squeezed momentum states and ideal GKP states) and the CV controlled-Z gate that entangles them.
In \cref{sec:clusterstates}, we decompose CVCSs and hybrid cluster states---including the special cases of GKP cluster states and another that we refer to as \emph{GKP-doped cluster states}, where the fraction of GKP states in the cluster is small~\cite{menicucci_fault-tolerant_2014, walshe_robust_2019}.
We introduce graphical descriptions for the subsystem-decomposed single-mode states and the various types of cluster state and use them to examine the entanglement structure of said states.
In \cref{sec:unzipping}, in a GKP-doped cluster-state setting, we show that the GKP states are the critical resource that ``unzips'' the remaining modes to reveal a logical-qubit cluster state and allowing the logical information to travel through the cluster state without acquiring noise from the gauge-mode entanglement.
Finally, in \cref{sec:fsqueez}, we describe how to extend this work to the case of physical resource states---\emph{i.e.},~finitely squeezed CVCSs (based on squeezed states instead of momentum eigenstates) and finitely-squeezed GKP states.

\section{Subsystem decomposition}
\label{sec:ssd}

Bosonic-mode subsystem decompositions are a way to decompose the Hilbert space of a CV mode into that of two virtual subsystems: (1)~a discrete, logical subsystem of dimension $d$ and (2)~a gauge mode---\emph{i.e.},~$\Hilbert_\CV \simeq \mathbb{C}^d \otimes \Hilbert_\CV'$~\cite{raynal_encoding_2012,  lau2017universalcontinuousvariable, marshall2019universalquantum, pantaleoni_modular_2020}. The SSD we consider here is based on a decomposition of the position operator in terms of an integer and a modular operator~\cite{aharonov_modular_1969}, inspired by---but not limited to---the periodic position wave functions of the GKP encoding~\cite{gottesman_encoding_2001}. We briefly outline the modular-position SSD, and we focus on the case of logical qubit subsystems ($d=2$). More details about the SSD can be found in Ref.~\cite{pantaleoni2021}.
The standard position basis for a bosonic mode comprises the set of eigenstates of the position operator $\op{q} = \frac{1}{\sqrt{2}}(\op{a} + \op{a}^\dagger)$. Each eigenstate $\ket{x}[q]$ is labeled by real eigenvalue  $x \in \mathbb{R}$ that can be written as a sum of three numbers,
\begin{align}
    x = \alpha \ell + 2 \alpha m + u
    \, ,
    \label{eq:spectrumdec}
\end{align}
where $\alpha$ is a fixed positive number known as the \emph{bin size}, $m \in \integers$ and $u \in [-\alpha/2,\alpha/2)$ are the \emph{gauge} quantum numbers, and $\ell \in \{0,1\}$ is the \emph{logical} quantum number. These quantum numbers result from two subsequent modular decompositions of $x$; details can be found in Refs.~\cite{pantaleoni_modular_2020, pantaleoni2021}.

The modular-position SSD is a change of basis from the position-quadrature basis to a tensor-product basis constructed by realizing that the quantum numbers in \cref{eq:spectrumdec} label two virtual subsystems:
\begin{align}
    \ket{x}[q]
        = \ket{\ell}[L]
            \otimes \ket{m, u}[G]
    \, .
    \label{eq:changeofbasis}
\end{align}
In the context of quantum computing, the logical subsystem---spanned by the $\lket{\ell}$ basis---plays the role of an encoded qubit, and the gauge subsystem---spanned by $\ket{m, u}[G]$---is a (virtual) bosonic mode~\cite{pantaleoni_modular_2020,pantaleoni2021}.

It may seem unusual that the gauge mode is specified by two separate quantum numbers (an integer $m$ and a real number in an interval $u$). A more standard description can be recovered by writing the ordinary position eigenvalue~$x_G$ of the gauge mode as
    \begin{align}
        x_G = \alpha m + u
        \, ,
    \end{align}
which is a modular decomposition of~$x_G$ with respect to $\alpha$.
The bin number $m$ labels the (centered) integer multiple of $\alpha$, and $u$ labels the (centered) fractional remainder. Since $m$ and $u$ describe a modular decomposition of the spectrum of the gauge-mode position operator~$\op q_G$, the states $\gket{m,u}$ comprise a basis for the gauge mode referred to as the \emph{partitioned-position} basis~\cite{pantaleoni_modular_2020}.

A different interpretation arises from noting that $m$ and $u$ are completely independent quantum numbers, which means they can be treated as acting on independent virtual subsystems~\cite{pantaleoni_modular_2020}---\emph{i.e.},~at least formally,
\begin{align}
\label{eq:rotordecomp}
    \gket{m,u} = \gket{m} \otimes \gket{u}\,.
\end{align}
This is a reflection of the fact that any CV mode's Hilbert space (in this case, that of the gauge mode) is isomorphic to that of two planar rotors~\cite{pantaleoni2021}. On the right-hand side of \cref{eq:rotordecomp}, $m$ labels the angular momentum states of one rotor, and $u$ labels the angular position of the other. (These two bases are mutually unbiased when applied to the same rotor.)

The isomorphism between the Hilbert space of a CV mode and that of two rotors exists even for the original CV mode, and one can apply a partitioned-position decomposition there, too, although we focus on the gauge mode here.\footnote{In fact, the modular-position subsystem decomposition can equivalently be interpreted as a decomposition of the original CV mode into two rotors followed by an encoding of a qubit into one of the rotors using the technique of Raynal~\emph{et al.}~\cite{raynal_encoding_2012}.}
In the sections that follow, we employ descriptions of the gauge subsystem both as a CV mode (in a partitioned-position basis) and also as two separate rotor subsystems. The latter has a tensor product description as
\begin{align}
    \ket{x}[q]
        = \ket{\ell}[L]
            \otimes \ket{m}[G] \otimes \ket{u}[G]
    \, ,
    \label{eq:changeofbasis2}
\end{align}
which is obtained by plugging \cref{eq:rotordecomp} into \cref{eq:changeofbasis}. While the physical system remains the same---a single physical mode---the two decompositions, Eqs.~\eqref{eq:changeofbasis} and~\eqref{eq:changeofbasis2}, have different conceptual meanings in terms of their treatment of the gauge subsystem---either as a single gauge mode or equivalently as two gauge rotors.
Throughout this work, we treat the gauge subsystem as a mode broken into two subsystems: the position bin-number subsystem (spanned by $\ket{m}[G]$) and the modular position subsystem (spanned by $\ket{u}[G]$). This is to keep the focus on the fact that the decomposition was originally motivated by binned homodyne detection on the original CV mode.

Several properties of the position-quadrature basis are inherited by the subsystem basis. First, orthogonality of the position-quadrature basis, $\braket{x}{x'}[q][q] = \delta(x-x')$, induces the orthogonality of the subsystem basis,
\begin{align}
    (
    \bra{\ell}[L]
        \bra{m,u}[G]
    )
        (
            \ket{\ell'}[L]
                \ket{m',u'}[G]
        )
            =
                \delta_{\ell \ell'}
                    \delta_{m m'}
                        \delta(u - u')
    \,.
\end{align}
Second, the completeness of position eigenstates, $\id_{\text{CV}}= \int dx \ketbra{x}{x}[q][q]$, means that the subsystem basis states are also complete over the mode,
\begin{align}
    \id_{\text{CV}}
        & =
            \sum_{\ell \in \{0,1\}}
            \hspace{-4pt}
            \ket{\ell}[L] \bra{\ell} [L] \otimes
                    \sum_{m \in \mathbb{Z}} \intalpha du
                        \ket{m,u}[G] \bra{m,u}[G]
                            \,.
    \label{eq:completeness}
\end{align}

From the subsystem decomposition, \cref{eq:changeofbasis2},
we construct diagonal logical and gauge-subsystem operators,
    \begin{subequations}
    \label{eq:decompops}
    \begin{align}
        \op{\ell} &\coloneqq \sum_{\ell \in \{0,1\}} \ell \ketbra{\ell}{\ell}[L][L] \, ,\\
        \op{m}_G &\coloneqq \sum_{m \in \mathbb{Z}} m \ketbra{m}{m}[G][G] \, ,\\
        \op{u}_G &\coloneqq \intalpha du \, u \ketbra{u}{u}[G][G]
        \, ,
    \end{align}
    \end{subequations}
which act only on the indicated subsystem---\emph{i.e.},~$\op \ell$ on the logical qubit, $\op m_G$ on the gauge bin-number subsystem, and $\op u_G$ on the gauge modular-position subsystem,
using the decomposition in \cref{eq:changeofbasis2}. We can recover the gauge position operator as
\begin{align}
    \op q_G = \alpha \op m_G + \op u_G\, .
\end{align}
Furthermore, using Eqs.~\eqref{eq:decompops}, along with \cref{eq:spectrumdec} and \cref{eq:completeness}, we can reconstruct the original CV-mode position operator,
\begin{align}
   \q = \alpha \l + 2 \alpha \mGop + \uGop
   \, .
   \label{eq:qsubsystemdecomp}
\end{align}
Operational descriptions of these operators can be found in Ref.~\cite{pantaleoni2021}.  We will heavily rely on \cref{eq:qsubsystemdecomp} when decomposing operators that are diagonal in $\q$.

\subsection{Decomposing momentum eigenstates and ideal Gottesman-Kitaev-Preskill states}
\label{sec:decsqueezedandGKP}

In their original formulation~\cite{menicucci_universal_2006}, ideal CV cluster states are composed of momentum eigenstates with eigenvalue 0, written~$\pket{0}$. The position wave function for these $0$-momentum states is constant, meaning that it has equal support in the position bins corresponding to the logical subsystem states $\lket{0}$ and $\lket{1}$. Moreover, the gauge-mode wave functions associated with these logical states are identical and are also constant, meaning that they describe another $0$-momentum state in the gauge mode. Both of these facts give a simple SSD,
    \label{eq:0momentumdef}
    \begin{align}
        \ket{0}[p]
        &= \lket{+} \otimes \ket{0}_{p,G}
        \\
            &=
    \lket{+}
    \otimes
    \sum_{m \in \mathbb{Z}} \gket{m}
    \otimes
    \intalpha du\, \gket{u}
    \, ,
    \end{align}
where, in the second line, we have decomposed the gauge-mode state as in \cref{eq:changeofbasis2}. More details can be found in Refs.~\cite{pantaleoni_modular_2020, pantaleoni2021}.
The key point is that a $0$-momentum state contains a logical $\lket{+}$ state with respect to the modular-position SSD.

The other type of CV resource state we consider is a square-lattice Gottesman-Kitaev-Preskill (GKP) state~\cite{gottesman_encoding_2001}, which encodes an error-correctable qubit into a CV mode using periodic wave functions. GKP states carry digital quantum information and interface seamlessly with the quantum-computational protocol realized by measuring CV cluster states with homodyne detection~\cite{walshe_2020}. In this setting, GKP states augment the computational power of CV cluster states by providing two things: (1)~encoded qubits and (2)~a means for error correction that ultimately allows for universality and fault tolerance~\cite{menicucci_fault-tolerant_2014, baragiola_all-gaussian_2019, mensen2020phasespace}.

While the $0$-momentum state, \cref{eq:0momentumdef}, is a single state, GKP states are a \emph{type} of state that encodes a qubit state in a two-dimensional subspace of a CV mode. The computational basis states, labeled by $j \in \{0,1\}$, are described by periodic superpositions of position eigenstates,
\begin{equation}
    \ket{j_{\GKP}}
        =
        \sum_{m \in \integers}
            \ket{\alpha (2m + j)}[q]
    \, .
    \label{eq:jgkp}
\end{equation}
An arbitrary qubit state, specified by amplitudes $c_0$ and $c_1$ satisfying $|c_0|^2 + |c_1|^2 = 1$, is
\begin{equation}
    \label{eq:arbGKP}
    \ket{\psi_{\GKP}}
        =
        c_0 \ket{0_{\GKP}}
            + c_1 \ket{1_{\GKP}}
    \, .
\end{equation}

The modular-position SSD of GKP states emerges straightforwardly after realizing that each position eigenstate appearing in the sum in \cref{eq:jgkp} decomposes simply as $ \qket{\alpha ( 2 m + j)} = \ket{j}[L] \otimes \ket{m,0}[G]$. This gives the SSD for each GKP computational basis state, $\ket{j_{\GKP}} = \ket{j}[L] \otimes \ket{+_\GKP}[G]$, where we used $\ket{+}[G] = \sum_{m}\ket{m,0}[G]$~\cite{pantaleoni2021}. By linearity, the arbitrary GKP state in \cref{eq:arbGKP} is a tensor-product state in the modular-position SSD,
    \begin{align}
        \ket{\psi_{\GKP}}
            &= \ket{\psi}[L] \otimes \ket{+}[G] \\
            &= \ket{\psi}[L] \otimes \sum_{m \in \mathbb{Z}} \ket{m}[G] \otimes \gket{u=0}
        \, ,
        \label{eq:GKParbSSD}
    \end{align}
with logical-subsystem state,
$
\ket{\psi}[L]
=
c_0 \ket{0}[L] + c_1 \ket{1}[L]
$.
We will later investigate GKP cluster states, where each mode is prepared in the specific GKP state that encodes a logical $\ket{+}$ state ($c_0 = c_1 = \frac{1}{\sqrt{2}}$),
    \begin{align}
    \label{eq:GKPplusdecomp}
        \ket{+_{\GKP}}
        &= \ket{+}[L] \otimes \sum_{m \in \mathbb{Z}} \ket{m}[G] \otimes \gket{u=0}
        \, .
    \end{align}
Note that the only difference between the SSD of $\ket{+_{\GKP}}$ that of the 0-momentum eigenstate in \cref{eq:0momentumdef} is that here the gauge-$u$ value is fixed to $u=0$.

\subsection{Decomposing CV controlled-$Z$ gates}
\label{subsec:CZdecompose}

To understand the entanglement structure of CV cluster states at the subsystem level, we decompose a specific two-mode unitary, the CV controlled-$Z$ gate. For $N$ modes, the decomposed Hilbert space is
$
(\Hilbert_\CV)^{\otimes N} \simeq (\complex^d \otimes \Hilbert_\CV') ^{\otimes N} = (\complex^d)^{\otimes N} \otimes (\Hilbert_\CV')^{\otimes N}
$. Operators that act on several of these modes decompose into pieces that may act exclusively in the logical subsystems, exclusively in the gauge modes, or across both logical and gauge subsystems. Below, we find that a decomposed CV controlled-$Z$ gate contains logical-only, gauge-only, and logical-gauge interactions terms. We will find that some of these terms can be ``turned off'' by tuning the interaction strength.

Canonical continuous-variable cluster states are built from $0$-momentum eigenstates coupled together by two-mode controlled-$Z$ gates~\cite{menicucci_universal_2006,gu_quantum_2009},
\begin{align}
    \CZ[g] \coloneqq e^{i g \q_1 \otimes \q_2 }
    \, ,
    \label{eq:CVcontrolledZ}
\end{align}
with interaction strength $g$ and numbered subscripts labeling the modes.
Using the subsystem decomposition of the position operator for both modes,
\begin{align}
    \q_i = \alpha \l_i + 2 \alpha \mGop[,i] + \uGop[,i]
    \,
\end{align}
for $i \in \{1,2\}$, we obtain
\begin{align}
 \label{eq:decomposedczgeneral}
     \CZ[g]
     &= \nonumber
     e^{i g \alpha^2 \l_{1} \otimes \l_{2} }
     e^{i g 4 \alpha^2 \mGop[,1] \otimes \mGop[,2] }
     e^{i g \uGop[,1] \otimes \uGop[,2] }
     \\ & \quad \times
       e^{i g 2 \alpha^2 (\l_{1} \otimes \mGop[,2] + \mGop[,1] \otimes \l_2) }
       e^{i g \alpha (\l_{1} \otimes \uGop[,2] + \uGop[,1] \otimes \l_2) }
     \nonumber \\ & \quad \times
       e^{i g 2 \alpha (\mGop[,1] \otimes \uGop[,2] + \mGop[,1] \otimes \uGop[,2]) }
     \, ,
\end{align}
where we have explicitly included the symbol $\otimes$ to emphasize the CV mode-wise tensor products inherited from \cref{eq:CVcontrolledZ}. Note that all of these exponentials commute.  The top line entangles subsystems of the same type across the two CV modes, the second line entangles the logical subsystem of one CV mode with the gauge mode of the other CV mode, and the final line couples the gauge modes of the two CV modes.

We focus on a $\CZ[g]$ gate ``tuned'' to have weight $g = \frac{\pi}{\alpha^2}$ commensurate with a chosen bin size $\alpha$. This weight simplifies \cref{eq:decomposedczgeneral}, since the discrete operators $\l$ and $\mGop$ have integer spectrum, so that $\text{exp}  (2\pi i\l_i \otimes \mGop[,j] ) = \exp (4 i \pi \mGop[,i] \otimes \mGop[,j])= \id$. The two-mode controlled-$Z$ gate decomposes into a product of logical-only, gauge-only, and logical-gauge interaction terms:
\begin{align}
    \label{eq:controlledzdecomposed}
    \CZone &= \CZL \CZg \CZint
    \, ,
\end{align}
each of which is an operator between different subsystems across the two CV modes.  The logical-only term is a controlled-$Z$ operator between the two qubit subsystems:
\begin{align}
    \CZL &\coloneqq  e^{i \pi \l_1 \otimes \l_2}
    \label{eq:logicalCZ}
    \, .
\end{align}
This operator is critical in uncovering hidden qubit cluster states within CVCSs, as it is the gate that provides the entangling interaction between qubits.

The gauge-only term is
\begin{align}
    \CZg & \coloneqq e^{ i \frac{\pi}{\alpha^2}
            \uGop[,1] \otimes \uGop[,2] }
            e^{ i \frac{2 \pi}{ \alpha } (\mGop[,1] \otimes \uGop[,2] + \uGop[,1] \otimes \mGop[,2]) }
    \, .
\end{align}
This operator does not contain a $\mGop[,1] \otimes \mGop[,2]$ coupling due to the chosen weight and the integer spectrum of the bin-number operators, as described above.
Finally, the interaction term,
\begin{align}
     \CZint & \coloneqq
    e^{
        i \frac{\pi}{\alpha}
        (
        \l_{1} \otimes \uGop[,2]
        +
        \uGop[,1] \otimes \l_{2}
        )
        }
    \, ,
    \label{eq:modphase&modrotation}
\end{align}
couples the logical subsystem of each CV mode to the modular gauge position of the other. The interaction operator can also be expressed as a product of two ``modular-shift'' operators generated by modular position,
$
      \exp\pqty{i \tfrac{\pi}{2\alpha} \uGop[,i]}
$,
and two controlled logical-$Z$ rotations,
$
\exp\pqty*{
    - i \tfrac{\pi}{2\alpha} \Z_{L,i} \otimes \uGop[,j \neq i]
}
$
, where the control is the modular gauge position of the other mode.

The $N$-mode generalization of the controlled-$Z$ operator, \cref{eq:controlledzdecomposed}, is
    \begin{align}
    \label{eq:multimodeCZ}
        \op{\text{C}}_Z[\mat{V}] \coloneqq \exp\pqty{ \frac{i}{2} \opvec{q}^\tp \mat{V} \opvec{q} }
        \, ,
    \end{align}
which describes position-position couplings between pairs of modes with weights specified by a real, $N\times N$, symmetric adjacency matrix $\mat{V}$, and where
    \begin{align}
        \opvec{q} \coloneqq \pqty*{\q_1,\dots,\q_N}^\tp
    \end{align}
is an $N$-dimensional column vector of position operators. The matrix transpose operation~$^\tp$ reshapes vectors but does not act at the operator level.

For concise notation, we introduce the matrix Kronecker product
\begin{align}
\label{eq:kronprod}
    \mat A \otimes \mat B\,,
\end{align}
which produces a block matrix by replacing each entry~$A_{jk}$ of~$\mat A$ with the matrix~$A_{jk} \mat B$. Importantly, $\mat A$ and $\mat B$ have no restrictions on their size or shape.

We now turn our attention to the SSD of $\op{\text{C}}_Z[\mat{V}]$, which induces decompositions of $\opvec{q}$ and the matrix $\mat{V}$ that we will make precise. Using \cref{eq:qsubsystemdecomp}, the position operator for mode~$j$ can be written as an inner product of the two column vectors
\begin{align}
    \vec \alpha
&
\coloneqq
    \begin{pmatrix}
        \alpha \\ 2\alpha \\ 1
    \end{pmatrix}
&
&\text{and}
&
    \opvec q_{\text{s},j}
&
\coloneqq
    \begin{pmatrix}
        \op \ell_j \\
        \op m_{G,j} \\
        \op u_{G,j}
    \end{pmatrix}
    \,,
\end{align}
representing, respectively, constant coefficients and a vector of subsystem operators for physical mode~$j$. Specifically,
\begin{align}
    \op q_j
=
    \vec \alpha^\tp \opvec q_{\text{s},j}\,.
\end{align}
We can collect all of the $\opvec q_{\text{s},j}$ into a $3N$-component column vector of subsystem operators,
\begin{align}
    \opvec q_\text{s}
&
\coloneqq
    (\opvec q_{\text{s},1}^\tp, \dotsc, \opvec q_{\text{s},N}^\tp)^\tp
\\*
&
=
    (\op \ell_1, \op m_{G,1}, \op u_{G,1}, \dotsc, \op \ell_N, \op m_{G,N}, \op u_{G,N})^\tp
    \, .
\end{align}
Note that the separation into 3-element blocks, one for each mode, is maintained. For later use, we define
$N$-component column vectors of subsystem operators of a single type:
\begin{subequations}
    \begin{align}
        \op{\Bell}
            & \coloneqq
            (\l_1, \dotsc , \l_N)^\tp
            \, ,\\*
        \opvec{m}_G
            & \coloneqq
            (\mGop[,1], \dotsc , \mGop[,N])^\tp
            \, ,\\*
        \opvec{u}_G
            & \coloneqq
            (\uGop[,1], \dotsc , \uGop[,N])^\tp
        \,.
    \end{align}
\end{subequations}
(The vector~$\opvec q_\text{s}$ may be obtained by interleaving these three vectors together.)

Using the Kronecker product, \cref{eq:kronprod}, we can write the whole vector of physical position operators as
\begin{align}
\label{eq:qvecdecomp}
    \opvec q
&
=
    (\mat \id \otimes \vec \alpha^\tp) \opvec q_\text{s}
    \,.
\end{align}
The Kronecker product in parentheses represents the appropriate ${N \times 3N}$ matrix of coefficients that does the job. Now, we can directly plug \cref{eq:qvecdecomp} into \cref{eq:multimodeCZ} to get
\begin{align}
    \label{eq:decomposedmultimodeCZ}
        \op{\text{C}}_Z[\mat{V}]
&
\coloneqq
    \exp\pqty{
    \frac{i}{2}
    \opvec q_\text{s}^\tp
    \mat V_\text{s}
    \opvec q_\text{s}
    }
    \, ,
\end{align}
where
\begin{align}
\label{eq:expandedcouplingmatrix}
    \mat V_\text{s}
&
\coloneqq
    (\mat \id \otimes \vec \alpha)
    \mat{V}
    (\mat \id \otimes \vec \alpha^\tp)
=
    \mat{V} \otimes (\vec \alpha \vec \alpha^\tp)
\\*
&
=
    \mat{V}
    \otimes
    \begin{pmatrix}
        \alpha^2 & 2\alpha^2 & \alpha \\
        2 \alpha^2 & 4\alpha^2 & 2\alpha \\
        \alpha & 2\alpha & 1 \\
    \end{pmatrix}
    \,.
\end{align}
This larger, ${3N \times 3N}$, adjacency matrix~$\mat V_\text{s}$ describes coupling between the subsystems and is given by the original adjacency matrix~$\mat V$ with each weight (entry)~$V_{ij}$ replaced by a network of connections described by the matrix~$V_{ij} \mat M_\text{s}$, which is known as a  \emph{matrix-valued weight} in the corresponding graph~\cite{flammia2009optical}. (This fact will be useful for visual illustration in the next section.)

For a subsystem decomposition based on bin size $\alpha$ for every mode, we consider the tuned multimode controlled-$Z$ gate with $\mat{V} = \tfrac{\pi}{\alpha^2} \mat{A}$, where $\mat{A}$ is a symmetric binary matrix (\emph{i.e.},~symmetric with entries of zero or one~\cite{menicucci_graphical_2011}) with diagonal elements set to $0$.\footnote{The more general setting where different modes have different bin sizes (and associated ``natural'' subsystem decompositions) is straightforward, but we do not dwell on it here.}
The adjacency matrix from \cref{eq:expandedcouplingmatrix}, written as
\begin{align}
\label{eq:expandedcouplingmatrixA}
    \mat{V}_\text{s}
        &=
        \frac{\pi}{\alpha^2}
            \mat{A} \otimes (\vec \alpha \vec \alpha^\tp)
        =
        \mat{A}
        \otimes
        \begin{pmatrix}
            \pi
                & 2\pi
                    &  \frac{\pi}{\alpha}
                        \\
            2\pi
                & 4\pi
                    & \frac{2 \pi}{\alpha}
                        \\
            \frac{\pi}{\alpha}
                & \frac{2 \pi}{\alpha}
                    & \frac{\pi}{\alpha^2}
                        \\
        \end{pmatrix}
        \, ,
\end{align}
reveals that the $\op{\ell}_{i} \otimes \op{m}_{G,j}$ and $\op{m}_{G,i} \otimes \op{m}_{G,j}$ terms in the multimode controlled-$Z$ gate, \cref{eq:decomposedmultimodeCZ}, generate trivial phases because the weights are integer multiples of $2\pi$, and the operator spectra are integers, just as in the two-mode case. Expanding the right-hand side of \cref{eq:decomposedmultimodeCZ} gives
\begin{align}
    \label{eq:multimodecontrolledzdecomposed}
    \CZ\big[\tfrac{\pi}{\alpha^2}\mat{A}\big] &=
    \CZL[ \mat{A}]
    \CZg [ \mat{A}]
    \CZint [ \mat{A}]
    \, ,
\end{align}
with
\begin{subequations}
\label{eq:multimodegeneralizedCZoperators}
    \begin{align}
        \CZL[ \mat{A}]
            & \coloneqq
                \exp\pqty{
                    \tfrac{i \pi }{2}
                        \op{\Bell}^\tp \mat{A} \op{\Bell}
                }
                \, ,
                \\
        \CZg [ \mat{A}]
            & \coloneqq
                \exp\pqty{
                    \tfrac{2 i \pi}{\alpha}
                        \opvec{m}_G ^\tp \mat{A} \opvec{u}_G
                            +
                                \tfrac{i \pi} {2 \alpha^2}
                                    \opvec{u}_G ^\tp \mat{A} \opvec{u}_G
                }
                \, ,
                \\
        \label{eq:multimodeinteractionoperator}
        \CZint [ \mat{A}]
            & \coloneqq
                \exp\pqty{
                    \tfrac{i \pi}{\alpha}
                        \op{\Bell}^\tp \mat{A} \opvec{u}_G
                }
        \,,
    \end{align}
\end{subequations}
noting that the coefficients in each operator vary, and we have used the facts that $\opvec{m}_G ^\tp \mat{A} \opvec{u}_G = \opvec{u}_G ^\tp \mat{A} \opvec{m}_G $ and $ \op{\Bell}^\tp \mat{A} \opvec{u}_G =   \opvec{u}_G^\tp \mat{A} \op{\Bell}$ (by the symmetry of~$\mat A$) to reduce the total number of terms.  This decomposition generalizes the tuned two-mode CV controlled-$Z$ in \cref{eq:controlledzdecomposed}, with all the same essential features. In particular, the interaction term can also be interpreted as a set of modular position shifts on each mode and a set of logical rotations around the $Z$-axis of each Bloch sphere that are controlled by modular gauge positions.

\section{Subsystem decomposition of cluster states}
\label{sec:clusterstates}

Continuous-variable cluster states can be used for universal CV quantum computing~\cite{lloyd_quantum_1999}---\emph{i.e.},~to implement any unitary of choice in the multimode Hilbert space $(\Hilbert_\CV)^{\otimes N}$. Fault tolerance is achievable if discrete-variable quantum information is encoded into the modes using a \emph{bosonic code} that enables error correction and if the initial error rate is low enough (which requires high enough squeezing)~\cite{menicucci_fault-tolerant_2014,bourassa2021blueprint,larsen2021faulttolerant}. A bosonic code is a prescription for encoding $d$-dimensional discrete quantum information into a mode, which can be done by choosing states in $\Hilbert_\CV$ that span a $d$-dimensional subspace~\cite{cochrane_macroscopically_1999,chuang_bosonic_1997,gottesman_encoding_2001,albert_performance_2018,grimsmo_quantum_2020}. The bosonic code that interfaces naturally the CV cluster state is the GKP code~\cite{gottesman_encoding_2001}, described briefly above in \cref{sec:decsqueezedandGKP}. CV cluster states serve as the substrate for fault-tolerant quantum computing when used in conjunction with the GKP encoding~\cite{menicucci_fault-tolerant_2014}.

We show here that every CV cluster state, which is composed of simple $0$-momentum eigenstates, \cref{eq:0momentumdef}, coupled together using Gaussian unitary gates encodes a ``hidden'' \emph{qubit} cluster state in the logical-subsystem degrees of freedom. However, this qubit cluster state is entangled with the gauge modes. Introducing GKP states into the cluster state---thus forming a \emph{hybrid cluster state}---changes the entanglement structure between the logical cluster state and the gauge modes. We show that a single GKP state can be used to ``unzip'' a linear CV cluster state, exposing and isolating the qubit cluster state hidden inside.

\subsection{ Decomposing CV cluster states}
\label{sec:idealcsdec}

The simplest non-trivial canonical CVCS is the two-mode state
\begin{align}
    \label{eq:twomodeCVCS}
    \ket{\text{CVCS}} =
    \CZ[g]
    \ket{0}[p,1] \otimes \ket{0}[p,2]
    \, ,
\end{align}
where $\CZ[g]$ is a CV controlled-$Z$ operator of weight $g$, \cref{eq:CVcontrolledZ}, and $\ket{0}[p]$ is a $0$-momentum eigenstate. We will decompose this state and introduce a graphical representation for this decomposition. This will be based on a hybrid of the graphical representation of ideal CV cluster states~\cite{menicucci_graphical_2011, zhang_local_2008, zhang_graphical_2008} and that of qubit-based graph states~\cite{hein_multiparty_2004}, suitably generalized to apply to the subsystems in our decomposition.

A $0$-momentum eigenstate, which we represent graphically as
\begin{equation}
\begin{aligned}
\begin{tikzpicture}[scale=1,yscale=1]
        \node[] (l1) at (-1,1)
            {$
                \ket{0}_{p} =
            $};
        \node[cvnode] (r3) at ( 0.08, 1.04) {};
        \node[] (r3) at ( 0.6, 1) {,};
            ;
\end{tikzpicture}
\label{diag:momeig}
\end{aligned}
\end{equation}
can be written as an unbiased superposition of position eigenstates as $\ket{0}[p] = (2\pi)^{-1/2} \int dx\, \qket{x}$ [the position wave function is the constant $(2\pi)^{-1/2}$].
Its subsystem decomposition [see \cref{eq:0momentumdef}] is
\begin{align}
    \ket{0}[p]
    &=
    \sum_{\ell  \in \{0,1\}} \lket{\ell}
    \otimes
    \sum_{m \in \mathbb{Z}} \gket{m}
    \otimes
    \intalpha du\, \gket{u}
    \label{eq:0momentumeig2}
    \, ,
\end{align}
recalling that $\lket{+} = \frac{1}{\sqrt{2}} (\lket{0} + \lket{1})$.  From this expression, we see that each factor of the decomposed $0$-momentum state---logical, gauge bin-number, and gauge modular position---is itself a constant superposition over a set of basis states for one of the subsystems. We represent each of these graphically as
\begin{subequations}
\begin{align}
\label{diag:momeigdec1}
    \sum_{\ell \in \{0,1\} }\ket{\ell}[L]
    &=
    \begin{aligned}
        \begin{tikzpicture}
            \node[logicalnode] () at (0,0) {};
        \end{tikzpicture}
    \end{aligned}
    \ , \\[5pt]
\label{diag:momeigdec2}
    \intalpha du\ \ket{u}[G]
    &=
    \begin{aligned}
        \begin{tikzpicture}
            \node[circlenode] () at (0,0) {};
        \end{tikzpicture}
    \end{aligned}
    \ , \\[10pt]
\label{diag:momeigdec3}
    \sum_{m \in \mathbb{Z}} \ket{m}[G]
    &=
    \begin{aligned}
        \begin{tikzpicture}
            \node[rectanglenode] () at (0,0) {};
        \end{tikzpicture}
    \end{aligned}
    \ .
\end{align}
\end{subequations}
Each symbol represents a specific state residing in a particular subsystem. The filled square and circle represent unbiased superpositions of all $\mGop$ and $\uGop$ eigenstates respectively, whereas the filled diamond represents an unbiased superposition of all $\l$ eigenstates (or, more simply, a logical $\lket{+}$ state). The coloring identifies the subsystem (black for undecomposed CV mode, blue for logical qubit, and red for gauge mode). With this, the graphical description of \cref{eq:0momentumeig2} is
\begin{equation}
    \begin{aligned}
        \label{eq:zeromomentumgraphicaldecomp}
        \begin{tikzpicture}[xscale=0.9,yscale=0.75]
                \node[cvnode]        (l1) at ( -2,2) {};
                \node[]              (l1) at ( -1,2) {$=$};
                \node[rectanglenode] (l1) at ( 0,1) {};
                \node[circlenode]    (l2) at ( 0,2) {};
                \node[logicalnode]   (l3) at ( 0,3) {};
                \node[]              ()   at (0.52,1.8) {.};
                    ;
        \end{tikzpicture}
    \end{aligned}
\end{equation}
Diagrams representing tensor products in multimode Hilbert spaces are obtained by simply appending more nodes. For example, the three-mode $0$-momentum eigenstate $\pket{0}^{\otimes3}$ is represented as
\begin{equation}
\begin{aligned}
\begin{tikzpicture}[xscale=0.9, yscale=0.75]
        \node[cvnode] (l1) at ( -5,2) {};
        \node[cvnode] (l1) at ( -4,2) {};
        \node[cvnode] (l1) at ( -3,2) {};
        \node[] (l1) at ( -2,2) {$=$};
        \node[rectanglenode] (l1) at ( -1,1) {};
        \node[circlenode] (l2) at ( -1,2) {};
        \node[logicalnode] (l3) at ( -1,3) {};
        \node[rectanglenode] (c1) at ( 0,1) {};
        \node[circlenode] (c2) at ( 0,2) {};
        \node[logicalnode] (c3) at ( 0,3) {};
        \node[rectanglenode] (r1) at ( 1,1) {};
        \node[circlenode] (r2) at ( 1,2) {};
        \node[logicalnode] (r3) at ( 1,3) {};
        \node[] () at ( 1.5,1.9 ) {.};
            ;
\end{tikzpicture}
\end{aligned}
\end{equation}
In this graphical description and those we present below, each column (on either side of the equal sign) represents a different CV mode. In the SSD on the right-hand side, each row represents a different subsystem type from \cref{eq:changeofbasis2}: the logical qubit in the top row and the gauge mode divided into bin-number and modular-position subsystems in the second and third rows, respectively. Note that these final rows can also be interpreted as representing rotor subsystems (as discussed in \cref{sec:ssd}).

Using the decomposition of $0$-momentum eigenstates, \cref{eq:0momentumeig2}, and of the $\CZ$ operator, \cref{eq:controlledzdecomposed}, we decompose the two-mode CV cluster state:
\begin{align}
    &
    \CZone\ket{0}[p,1] \otimes \ket{0}[p,2]
    =
    \CZint
    \big( \lket{\text{CS}} \otimes \gket{\Phi} \big)
    \, ,
    \label{eq:decomposed2modecs}
\end{align}
where
\begin{align}
    \label{eq:twoqubitCS}
    \lket{\text{CS}}
        \coloneqq\CZL
        \ket{+}[L,1]
        \ket{+}[L,2]
    \end{align}
is a logical two-qubit cluster state, which is entangled with the gauge-mode state
    \begin{align}
    \gket{\Phi} \coloneqq
    \CZg
    \ket{0}[p,G,1]
    \ket{0}[p,G,2]
    \end{align}
via the interaction $\CZint$. The decomposition reveals that an ordinary qubit cluster state is ``hidden inside'' the CV cluster state~\cite{pantaleoni_modular_2020}, but it remains entangled with the gauge modes. In what follows, we represent this fact graphically. Note that we have suppressed the tensor-product notation $\otimes$ for brevity, a convention we continue with henceforth.

The graphical representation of a simple two-mode CV cluster state with weight $g$ is~\cite{gu_quantum_2009,menicucci_graphical_2011}
\begin{equation}
    \begin{aligned}
    \label{eq:twomodeCVCSgraphical}
    \begin{tikzpicture}[]
            \node[cvnode] (cv1) at ( 1.5,0 ) {};
            \node[cvnode] (cv2) at ( 3,0 ) {};
            \node[] () at ( 3.5, -0.1 ) {.};
            \node[] () at ( 2.25, 0.3 ) {$g$};
            \node[] (cv3) at ( -1,0) {$\CZ[g] \ket{0}[p,1] \ket{0}[p,2] \; = $};
            \path
                (cv1) edge[cvline] (cv2)
                ;
    \end{tikzpicture}
    \end{aligned}
\end{equation}
Filled black circles still represent $0$-momentum eigenstates. The line between the nodes is a CV controlled-$Z$ gate of weight $g$.
Since the decomposition of the two-mode CVCS in \cref{eq:decomposed2modecs} relies on the specific weight $g = \frac{\pi}{\alpha^2}$, we now consider cluster states generated by these ``properly tuned'' CV controlled-$Z$ gates.
\begin{figure*}[t]
    \centering
    \includegraphics[width=0.9\textwidth]{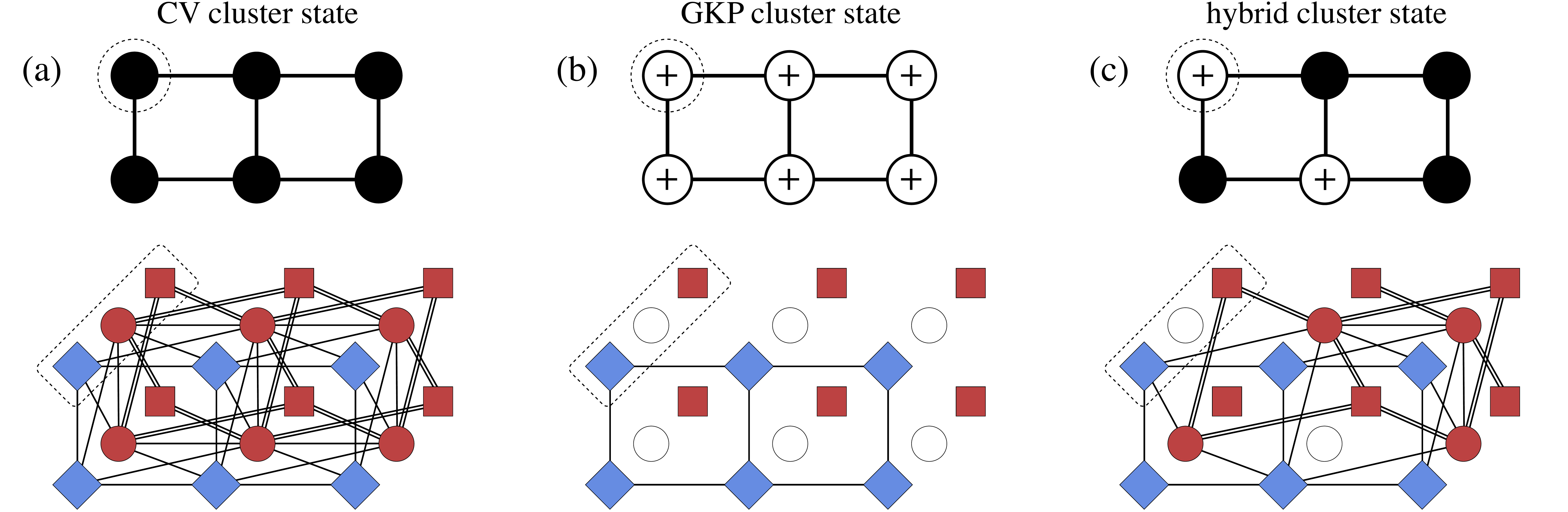}
    \caption{
    Graphical depictions of three types of 6-mode cluster state and their subsystem decompositions. The top row shows cluster states at the level of the CV modes, and the bottom row shows the same states in the SSD. For reference, a single mode and its SSD are circled by dotted lines in each subfigure. (a) CV cluster state, \cref{eq:logicalqubitclusterstateA}. (b) GKP cluster state, \cref{eq:decomposedcgkpvcs}. (c) Hybrid cluster state, \cref{eq:hybridCSgeneral}. The logical-qubit cluster state (blue diamonds) is the same for (a), (b), and (c); what differ are the state over the gauge modes and the logical-gauge entanglement.
    }
    \label{fig:graphicaldecompositions}
\end{figure*}
To represent the decomposed two-mode CVCS, we introduce graphical descriptions for various pairs of entangled subsystems. Of primary interest to us is the hidden logical-qubit cluster state, \cref{eq:twoqubitCS}, which is represented by the following diagram:
\begin{equation}
\begin{aligned}
\begin{tikzpicture}[scale=0.75,yscale=1]
        \node[] () at (-3,0) {$\lket{\text{CS}}$};
        \node[] () at ( -2,0) {$=$};
        \node[logicalnode] (l) at ( -1,0) {};
        \node[logicalnode] (r) at ( 1,0) {};
        \node[] () at ( 1.5,0) {.};
        \path
            (l) edge[] (r)
            ;
\end{tikzpicture}
\end{aligned}
\end{equation}
This is the type of diagram used in cluster-state literature to describe qubit-cluster states~\cite{nielsen_optical_2004}. Similarly, we define graphical representations for interactions between unbiased states across different subsystem types:
\begin{align}
     \label{eq:twosystemgraphical}
     e^{
        i\pi ( \l \otimes \tfrac{\uGop}{\alpha} )
       }
    \ket{+}[L] \int du \ket{u}[G]
    =
    \begin{aligned}
        \begin{tikzpicture}[yscale=0.8,xscale=0.9]
            \node[logicalnode] (l1) at ( -0.9,1) {};
            \node[circlenode] (r1) at ( 0.1,1) {};
            \path
                (l1) edge[] (r1) ;
        \end{tikzpicture}
    \end{aligned}
    \,,
    \\
    \label{eq:twosystemgraphical2}
    e^{
        i\pi ( \frac{\uGop}{\alpha} \otimes \mGop )
      }
    \int du \ket{u}[G] \sum_{m}\ket{m}[G]
    =
    \begin{aligned}
        \begin{tikzpicture}[yscale=0.8,xscale=0.9]
            \node[circlenode] (l1) at ( -0.9,1) {};
            \node[rectanglenode] (r1) at ( 0.1,1) {};
            \path
                (l1) edge[] (r1) ;
        \end{tikzpicture}
    \end{aligned}
    \,.
\end{align}
As in \crefrange{diag:momeigdec1}{diag:momeigdec3}, diamonds are unbiased logical states, red circles are unbiased gauge modular position states, and squares are unbiased bin-number states. Lines are interactions of the type $\exp(i \pi \ \cdot \ \otimes \ \cdot )$, where the placeholders ``$\cdot$'' are filled in depending on the interacting subsystems. In particular, we have chosen to associate a circle node with $ \uGop / \alpha $, a square node with $ \mGop $, and a logical node with $\l$. With these associations, any line connecting subsystem-decomposed nodes is a gate with strength $\pi$.
Using these definitions, we can write down the graphical representation of the decomposed two-mode CVCS in \cref{eq:twomodeCVCSgraphical} as
\begin{equation}
\begin{aligned}
\label{twomodeCVCSgraphicaldecomp}
\begin{tikzpicture}[xscale=0.9,yscale=0.75]
        \node[cvnode] (cv1) at ( -4.8,2) {};
        \node[cvnode] (cv2) at ( -2.8,2) {};
        \node[] () at ( -2,2) {$=$};
        \node[rectanglenode] (l1) at ( -1,1) {};
        \node[circlenode] (l2) at ( -1,2) {};
        \node[logicalnode] (l3) at ( -1,3) {};
        \node[rectanglenode] (r1) at ( 1,1) {};
        \node[circlenode] (r2) at ( 1,2) {};
        \node[logicalnode] (r3) at ( 1,3) {};
        \node[] () at ( 1.5,1.9) {,};
        \node[] () at ( -3.75, 2.4 ) {$\pi \alpha^{-2}$};
        \path
            (cv1) edge[cvline] (cv2)
            (l1) edge[doubleline] (r2)
            (l2) edge[doubleline] (r1)
            (l2) edge[] (r3)
            (l2) edge[] (r2)
            (l3) edge[] (r2)
            (l3) edge[] (r3)
            ;
\end{tikzpicture}
\end{aligned}
\end{equation}
where the double lines indicate two instances of the same gate. From this description, it is again apparent that a two-mode CVCS includes a two-qubit cluster state entangled with the gauge modes---specifically, the gauge modular position.

The above two-mode description extends in a straightforward way to $N$-mode CV cluster states. Applying the $ \op{\text{C}}_Z \big[\tfrac{\pi}{\alpha^2}\mat{A}\big]$ operator to $N$ modes prepared in $\pket{0}^{\otimes N}$ generates an $N$-mode CV cluster state. Using the decomposition in \cref{eq:multimodecontrolledzdecomposed},
\begin{align}
    \ket{
        \text{CVCS}_{
            \frac{\pi}{\alpha^2}
                \mat{A}
        }
    }
    & \coloneqq
        \op{\text{C}}_Z
        \big[
            \tfrac{\pi}{\alpha^2}
                \mat{A}
        \big]
            \pket{0}^{\otimes N}
    \\
    &=
        \CZint [ \mat{A}]
        \big(
            \lket{
                \text{CS}_{\mat{A}}
            }
                \otimes \gket{\Phi_\mat{A}}
        \big)
    \,,
    \label{eq:logicalqubitclusterstateA}
\end{align}
where the states in the second line are defined below.
Importantly, the logical-only part of the decomposed controlled-$Z$ operator in \cref{eq:multimodecontrolledzdecomposed} contains the binary adjacency matrix $\mat{A}$ that describes the entanglement structure in the hidden logical-qubit cluster state,
\begin{align}
    \lket{
        \text{CS}_{\mat{A}}
    }
        \coloneqq
            \CZL[ \mat{A} ]
                \lket{+} ^{\otimes N}
    \,.
\label{eq:logicalqubitclusterstate}
\end{align}
Through the interaction operator $\CZint [ \mat{A}]$, this cluster state is entangled to the $u_G$-components of the gauge-mode state
    \begin{align}
        \gket{ \Phi_{\mat{A}} } \coloneqq \CZg [ \mat{A}] \ket{0}_{p,G}^{\otimes N}
        \, .
    \end{align}
For reference, the two-mode CV cluster state in \cref{eq:decomposed2modecs} is described by $\mat{V} = \frac{\pi}{\alpha^2} \mat{A}$, with
    \begin{align}
    \label{eq:twomodeadjacencymatrix}
        \mat{A} =
        \begin{pmatrix}
            0 & 1 \\
            1 & 0
        \end{pmatrix}
        \, .
    \end{align}
Entanglement between the logical and gauge subsystems precludes the use of these qubit cluster states directly, so
it is necessary to devise a strategy to combat potential logical decoherence caused by the interaction operator, hence freeing up the logical-qubit CV cluster state. In \cref{sec:unzipping}, we present a method to liberate the logical cluster state from the gauge modes using supplemental GKP states.

A graphical depiction of a CV cluster state defined on 6 modes is shown in Fig.~\ref{fig:graphicaldecompositions}(a).
The connections between nodes in graphical representations of CV cluster states, including the two-mode state in \cref{twomodeCVCSgraphicaldecomp}, are those given by the matrix $\mat{V}_\text{s}$ in \cref{eq:expandedcouplingmatrix}.

\subsection{Decomposing GKP cluster states}

It is not strictly necessary that interactions between GKP-encoded qubits are mediated by continuous-variable cluster states, although designing schemes that use both can present a number of advantages---see Refs.~\cite{bourassa2021blueprint,larsen2021faulttolerant} for two examples of the GKP-CVCS synergy in action. In fact, a number of schemes have been proposed where multiple GKP codewords interact either in a measurement-based fashion with no need for CVCSs or with direct control over two-mode interactions~\cite{noh2020fault,vuillot_quantum_2019,fukui_high-threshold_2019,fukui_high-threshold_2018,fukui_analog_2017, yamasaki2020polylogoverhead, hanggli_enhanced_2020, noh2021low}.

We focus here on \emph{GKP cluster states}, where every node in the cluster is prepared in a $\ket{+_\GKP}$ state (rather than a $0$-momentum state as for CVCSs). Over $N$ modes, a GKP cluster state is given by
\begin{align}
        \ket{ \text{GKPCS}_{\mat{V}} } &\coloneqq \CZ[ \mat{V}] \ket{+_\GKP}^{\otimes N}
    \, ,
\end{align}
using the multimode controlled-$Z$ operator in \cref{eq:multimodeCZ}.
GKP cluster states are directly compatible with measurement-based quantum computing protocols originally designed for CV cluster states~\cite{menicucci_universal_2006}, with the added advantage that they have baked-in potential for error correction~\cite{walshe_2020}.
The foundation for the utility of GKP cluster states is that they contain a logical-qubit cluster state, but unlike the case of CV cluster states, \cref{eq:logicalqubitclusterstate}, this qubit cluster is not entangled with the gauge modes and is thus ready to use.

Using the SSD of $\ket{+_\GKP}$, \cref{eq:GKPplusdecomp}, and an appropriately tuned multimode controlled-$Z$ operator (${\mat{V} = \frac{\pi}{\alpha^2} \mat{A}}$), \cref{eq:multimodecontrolledzdecomposed}, we find the SSD for a GKP cluster state,
\begin{align}
    \ket{\text{GKPCS}_{\tfrac{\pi}{\alpha^2}\mat{A}}}
        &=
            \lket{
                \text{CS}_{\mat{A}}
            } \otimes
                \ket{+_{\GKP}}[G] ^{\otimes N}
    \label{eq:decomposedcgkpvcs}
    \, .
\end{align}
No gauge-mode or interaction operators, such as the ones defined in \cref{eq:multimodegeneralizedCZoperators}, appear in the subsystem-decomposed form---more precisely, these operators act trivially as the identity operator due to the fact that $u_G = 0$ for GKP states. This decouples the logical-qubit cluster state $\ket{\text{CS}_{\mat{A}}}[L]$, \cref{eq:logicalqubitclusterstate}, from the gauge modes, which are themselves in a tensor-product state.

GKP cluster states have simple graphical representations. We begin with the graphical depiction of an ideal GKP state $\ket{+_\GKP}$ on a single CV mode,
\begin{equation}
\begin{aligned}
\begin{tikzpicture}[]
        \node[] (l1) at (-1,0)
            {$ \ket{+_\GKP} \ = $};
        \node[emptycvnode] (cv) at ( 0.4,0) {};
        \node[font=\large] () at ( 0.4,0) {$\boldsymbol{+}$};
        \node[] (r3) at ( 1,0) {.};
            ;
\end{tikzpicture}
\label{diag:gaugemomeig}
\end{aligned}
\end{equation}
In the subsystem decomposition, the difference between a $0$-momentum state and an ideal $\ket{+_\GKP}$ state lies only in the gauge mode. Namely, for $0$-momentum states, it is an unbiased superposition of $u_G$ eigenstates $\intalpha \ du \gket{u}$, while for $\ket{+_\GKP}$ states, the gauge mode is in an eigenstate of $\op{u}_G$ with eigenvalue 0, $\gket{u = 0}$. The former is represented graphically as a filled red circle, and we represent the latter as an empty circle,
\begin{equation}
\begin{aligned}
\begin{tikzpicture}[scale=1,yscale=1]
        \node[] (l2) at (-1,2)
            {$
                \ket{u=0}[G] \; = \;
            $};
        \node[emptycirclenode] (r2) at (0.35,2) {};
        \node[] (r2) at (0.7,1.9) {,};
\end{tikzpicture}
\label{diag:ugaugezero}
\end{aligned}
\end{equation}
so that the SSD of a GKP $\ket{+}$ state is represented graphically as
\begin{equation}
\begin{aligned}
\begin{tikzpicture}[xscale=0.9,yscale=0.75]
        \node[emptycvnode] (cv) at           (-2,2) {};
        \node[font=\large]
                          () at         (-2,2)
                            {$\boldsymbol{+}$};
        \node[] () at                   (-1,2) {$=$};
        \node[rectanglenode] (r1) at    ( 0,1) {};
        \node[emptycirclenode] (r1) at  ( 0,2) {};
        \node[logicalnode] (r1) at      ( 0,3) {};
        \node[] (r1) at                 (0.5,1.9) {.};
            ;
\end{tikzpicture}
\end{aligned}
\end{equation}
Compared to the graphical decomposition of the $0$-momentum state, \cref{eq:zeromomentumgraphicaldecomp}, the only difference is in the modular position subsystem---the color of the circle.

A two-mode GKP cluster state, \cref{eq:decomposedcgkpvcs}, with binary adjacency matrix $\mat{A}$ in \cref{eq:twomodeadjacencymatrix}, is
    \begin{align}
    \label{eq:twomodeGKPCS}
        \ket{\text{GKPCS}} = \lket{\text{CS}} \ket{+_\GKP}_{G,1} \ket{+_\GKP}_{G,2}
        \, ,
    \end{align}
with graphical depiction
\begin{equation}
\begin{aligned}
\begin{tikzpicture}[xscale=0.9,yscale=0.75]
    \node[emptycvnode] (cv1) at                (-5,2) {};
    \node[font=\large] () at   (-5,2)
         {$\boldsymbol{+}$};
    \node[emptycvnode] (cv2) at                (-3,2) {};
    \node[font=\large] () at   (-3,2)
         {$\boldsymbol{+}$};
    \node[] () at (-2,2) {$=$};
        \node[rectanglenode] (l1) at      (-1,1) {};
        \node[emptycirclenode] (l2) at    (-1,2) {};
        \node[rectanglenode] (r1) at      ( 1,1) {};
        \node[emptycirclenode] (r2) at    ( 1,2) {};
        \node[logicalnode] (l3) at        (-1,3) {};
        \node[logicalnode] (r3) at        ( 1,3) {};
        \node[] (l2) at                   (1.5,1.9)
            {.};
        \path
            (l3) edge[] (r3)
            (cv1) edge[cvline] (cv2)
            ;
\end{tikzpicture}
\end{aligned}
\end{equation}
From this description, it is clear that \emph{only} the logical-qubit subsystems are entangled by the controlled-$Z$ gate, which puts them in a two-qubit cluster state, \cref{eq:twoqubitCS}, unentangled with the gauge modes. A 6-mode example is shown in Fig.~\ref{fig:graphicaldecompositions}(b).

At this stage, the reader may have noticed that none of the decompositions we presented require explicit use of the conjugate quadrature operator $\op{p}$. This is, in fact, a feature of ideal cluster states that considerably simplifies the present analysis. The subsystem decomposition of position shifts (that is, the exponentiated version of $\op{p}$) can be found in Ref.~\cite{pantaleoni2021}. Formulas for expressing the momentum quadrature in terms of modular and integer operators, along with their commutation relations with modular and integer position, can be found elsewhere~\cite{aharonov_modular_1969,englert_periodic_2006}. Note that the definition of a set of conjugate logical and gauge operators analogous to $\l$, $\mGop$ and $\uGop$ is a more delicate problem which we leave to future work.

\subsection{Supplementing CV cluster states with GKP states: hybrid cluster states}
\label{sec:supplementing}

Several proposals for fault-tolerant quantum computing use Gaussian squeezed states and GKP states together~\cite{menicucci_fault-tolerant_2014, walshe_robust_2019}. In these references, the squeezed states are used to generate a CVCS---the resource that enables computation---while the GKP states are used (1)~as carriers of logical information and (2)~for error correction to help undo noise that accumulates during computation. In these studies, the GKP states are sparsely distributed, and we refer to such entangled resource states as \emph{GKP-doped cluster states}. These are a subset of more general \emph{hybrid cluster states}, for which some fraction of the total nodes are $\ket{+_\GKP}$ states, and the remainder are $0$-momentum states.
Proposals for all-optical GKP sources are not deterministic, so hybrid cluster states are likely to be prepared using some combination of deterministic squeezed states and probabilistic GKP states. The proposal from Bourassa \emph{et al.}~\cite{bourassa2021blueprint} shows that hybrid cluster states arranged in 3D Raussendorf-Harrington-Goyal lattice~\cite{raussendorf_fault-tolerant_2006} can be fault tolerant if the fraction of randomly distributed GKP states is high enough (at least 76.4\%), although smaller fractions may be possible with better decoders.

Here, we introduce and provide the subsystem decomposition for ideal hybrid cluster states produced when multimode controlled-$Z$ gates are used to couple $0$-momentum states and ideal $\ket{+_\GKP}$ states. An $N$-mode hybrid cluster state takes the general form,
   \begin{align}
   \label{eq:hybridCSgeneral}
        \ket{ \text{hybridCS}_{\mat{V}} } &\coloneqq \CZ[ \mat{V}] \bigotimes_{i=1}^N \ket{\psi_i}
    \,,
    \end{align}
where $\ket{\psi_i} \in \{ \ket{0}_{p,i} , \ket{+_\GKP}_i \}$ is the CV state in mode~$i$. This form allows for mixing and matching of $\pket{0}$ and $\ket {+_\GKP}$ at various places in the cluster state.

To illustrate hybrid cluster states, we start with the simplest two-mode situation, where one mode is prepared in a $0$-momentum eigenstate and the other in the ideal GKP state~$\ket{+_{\GKP}}$.
Applying the properly tuned multimode controlled-$Z$ operator ($\mat{V} = \frac{\pi}{\alpha^2} \mat{A}$), \cref{eq:multimodecontrolledzdecomposed}, gives
\begin{align}
   &\CZ[\tfrac{\pi}{\alpha^2}\mat{A}] \ket{0}[p,1] \ket{+_{\GKP}}[2]
     =
    e^{ i \frac{\pi}{\alpha}
    \uGop[,1] \otimes \l_{2} }
    \lket{\text{CS}} \gket{\Phi}
    \label{eq:hybridstate}
    \, .
\end{align}
where $\lket{\text{CS}}$ is the two-qubit logical cluster state in \cref{eq:twoqubitCS}, and the gauge mode state is
    \begin{align}
        \gket{\Phi} = e^{ i \frac{\pi}{\alpha} \uGop[,1]  \mGop[,2] } \ket{0}[p,G,1] \ket{+_{\GKP}}[G,2]
        \, .
    \end{align}
Since ideal GKP states have $u_G = 0$, the coupling is asymmetric across the modes' subsystems. Recall the operator
\begin{align}
    \rotmat (\theta)
        =
        \exp\pqty*{
            - i \theta \op Z _L /2
        }
        ,
\end{align}
which rotates the Bloch vector of the state by an angle~$\theta$ around the computational-basis~($Z$) axis. Formally replacing the angle~$\theta$ with~$\tfrac \pi \alpha \uGop[,1]$ in this operator,
we can write the interaction operator in \cref{eq:hybridstate} as
\begin{align}
\label{eq:controlledrot}
    \exp\pqty{
        i \frac{\pi}{\alpha}
            \uGop[,1]
                \otimes \l_{2}
    }
    &=
    \exp\pqty{
        i \frac{\pi}{2\alpha}
            \uGop[,1]
    }
        \rotop_{L,2}
            \pqty{\frac{\pi}{\alpha} \uGop[,1] }
\,,
\end{align}
describing a simultaneous phasing of the modular-position subsystem of the first mode and a rotation of the logical qubit of the second mode controlled on~$\op u_G$ of the first mode~\cite{pantaleoni2021}.

Hybrid cluster states over more modes are constructed straightforwardly using properly tuned interactions, ($\mat{V} = \frac{\pi}{\alpha^2} \mat{A}$) in \cref{eq:hybridstate}, along with the two-mode decompositions for the three possible combinations of CV state at each node: momentum eigenstate-momentum eigenstate [\cref{eq:decomposed2modecs}], momentum eigenstate-GKP [\cref{eq:hybridstate}], and GKP-GKP [\cref{eq:twomodeGKPCS}].

A graphical depiction of the 2-mode hybrid cluster state in \cref{eq:hybridstate} is given by
\begin{equation}
\begin{aligned}
\begin{tikzpicture}[xscale=0.9,yscale=0.75]
        \node[cvnode] (cv1) at (-5,2) {};
        \node[emptycvnode] (cv2) at (-3,2) {};
        \node[font=\large] (cv2l) at (-3,2)
            {+};
        \node[] () at (-2,2) {$=$};
        \node[rectanglenode] (l1) at (-1,1) {};
        \node[circlenode] (l2) at (-1,2) {};
        \node[rectanglenode] (r1) at ( 1,1) {};
        \node[emptycirclenode] (r2) at ( 1,2) {};
        \node[logicalnode] (l3) at (-1,3) {};
        \node[logicalnode] (r3) at ( 1,3) {};
        \node[] () at ( 1.5,1.9) {,};
        \path
            (cv1) edge[cvline] (cv2)
            (r1) edge[doubleline] (l2)
            (l2) edge[] (l2)
            (l2) edge[] (r3)
            (l3) edge[] (r3)
            ;
\end{tikzpicture}
\end{aligned}
\end{equation}
where the asymmetric interaction gives rise to the line between the logical qubit of mode 2 and the gauge modular position of mode 1.
The graphical depiction of a 6-mode example is given in Fig.~\ref{fig:graphicaldecompositions}(c).

\newcommand{\GKPsmall}{
        \resizebox{10pt}{3pt}{\text{GKP}}
}

In fact, the entanglement structure of a hybrid cluster state simplifies for \emph{any} GKP-encoded logical state where
$
\ket{\psi_{\GKP}} = \ket{ \psi }[L] \otimes \sum_m\ket{m}[G] \otimes \ket{u =0}[G]$ [\cref{eq:GKParbSSD}]. In terms of subsystem diagrams, we have
\begin{equation}
\begin{aligned}
\begin{tikzpicture}[scale=0.9,yscale=0.75]
        \node[cvnode] (cv1) at (-5,2) {};
        \node[emptycvnode] (cv2) at (-3,2) {};
        \node[font=\footnotesize] () at (-3,2) {
        $\psi_{\GKPsmall}$
        };
        \node[] () at (-2,2) {$=$};
        \node[rectanglenode] (l1) at (-1,1) {};
        \node[circlenode] (l2) at (-1,2) {};
        \node[rectanglenode] (r1) at (1,1) {};
        \node[emptycirclenode] (r2) at (1,2) {};
        \node[logicalnode] (l3) at (-1,3) {};
        \node[emptylogicalnode] (r3) at (1,3) {};
        \node[] () at (1,3) {\footnotesize $\psi$};
        \node[] () at (1.5,1.9) {,};
        \path
            (cv1) edge[cvline] (cv2)
            (r1) edge[doubleline] (l2)
            (l2) edge[] (l2)
            (l2) edge[] (r3)
            (l3) edge[] (r3)
            ;
\end{tikzpicture}
\end{aligned}
\end{equation}
where the $\psi$-diamond represents the logical qubit in state $\ket{\psi}[L]$.

\section{Unzipping a CVCS by teleporting GKP states}
\label{sec:unzipping}
While a continuous-variable cluster state~(CVCS) is not explicitly encoded (in the sense that it is not \emph{defined} as a codeword for any bosonic code), we have shown that, by decomposing CV Hilbert space, we can endow these CV states with a logical-subsystem qubit. This is accomplished by simply decomposing the state and using the information within the logical subsystem.

When we use this idea on CVCSs, we run into the problem that, even though we are able to recognize a useful feature hidden within the CV cluster state---a logical-qubit cluster state, as illustrated in \cref{eq:decomposed2modecs} and shown in Fig.~\ref{fig:graphicaldecompositions}(a)---it is unclear how to gain direct access to it due to entanglement with the gauge modes. Furthermore, we know that the GKP encoding dovetails particularly well with CVCSs~\cite{menicucci_fault-tolerant_2014} since homodyne detection enables all Gaussian unitaries, and this includes all GKP Cliffords. This suggests there should be some connection between the GKP encoding and CVCSs, and the SSD---through Fig.~\ref{fig:graphicaldecompositions}---provides some clues for what this connection is.

From Fig.~\ref{fig:graphicaldecompositions}(b), we know that a GKP cluster state is just a logical-qubit cluster state that is fully disconnected from its gauge subsystems. So it is no surprise that the measurement sequences required to implement gates on this cluster state work as they should since we are directly implementing qubit-level measurement-based quantum computing on a qubit cluster state using GKP as the dictionary to implement it bosonically. What is surprising is that modes that are used merely to implement gates---not to carry quantum information or correct errors---need not be GKP states! In fact, in the ideal case, the hybrid cluster state in Fig.~\ref{fig:graphicaldecompositions}(c) works just as well for processing GKP quantum information as the state in Fig.~\ref{fig:graphicaldecompositions}(b). And if we take it to the extreme, the fully CV cluster state in Fig.~\ref{fig:graphicaldecompositions}(a) also serves perfectly well as a resource for GKP measurement-based quantum computation as long as we can teleport in the initial GKP-encoded quantum information. While the behavior with respect to errors will be different between (a), (b), and~(c), in the limit of high-quality states, these are all just as good. This is true despite the fact that (a) has lots of gauge-mode entanglement, while (b) has none.

This curious fact leads to an important question that the SSD can help us to answer: \emph{Why do CVCSs serve so well as logical-qubit cluster states when they have so much entanglement with the gauge modes?} One would expect this gauge-logical entanglement to spoil the simplicity of using these states to process GKP quantum information, but for some reason it does not. The answer to this question lies in the subtle way that the subsystem-decomposed structure of GKP-encoded qubits ``unzips'' the cluster state---disconnecting the logical information from the adjacent gauge modes---with every teleportation step. Let us examine this further with a simple example.

We will analyze a one-dimensional CVCS, also known as a CV quantum wire~\cite{alexander_noise_2014}, with a single logical GKP state attached via CV controlled-$Z$ (with an appropriate weight~$g$). This hybrid cluster state, with logical input~$\ket{\psi}[L]$, looks like this:
\begin{equation}\label{tikz:CVQW}
\begin{tikzpicture}[xscale=0.8,yscale=0.75]
        \node[cvnode] (cv1) at (-9,2) {};
        \node[cvnode] (cv2) at (-8,2) {};
        \node[cvnode] (cv3) at (-7,2) {};
        \node[cvnode] (cv4) at (-6,2) {};
        \node[emptycvnode] (cv5) at (-5,2) {};
        \node[] () at (-5,2) {$\psi_{\GKPsmall}$};
        \node[] () at (-4.1,2) {$=$};
        \node[rectanglenode] (lll1) at    (-3.3,1) {};
        \node[circlenode] (lll2) at       (-3.3,2) {};
        \node[logicalnode] (lll3) at      (-3.3,3) {};
        \node[rectanglenode] (ll1) at     (-2.3,1) {};
        \node[circlenode] (ll2) at        (-2.3,2) {};
        \node[logicalnode] (ll3) at       (-2.3,3) {};
        \node[rectanglenode] (l1) at      (-1.3,1) {};
        \node[circlenode] (l2) at         (-1.3,2) {};
        \node[logicalnode] (l3) at        (-1.3,3) {};
        \node[rectanglenode] (r1) at      (-0.3,1) {};
        \node[circlenode] (r2) at         (-0.3,2) {};
        \node[logicalnode] (r3) at        (-0.3,3) {};
        \node[rectanglenode] (rr1) at     ( 0.7,1) {};
        \node[emptycirclenode] (rr2) at   ( 0.7,2) {};
        \node[emptylogicalnode] (rr3) at       ( 0.7,3) {};
        \node[] () at           ( 0.7,3)
            {{\footnotesize $\psi$}};
        \node[] () at                     (1.2,1.9) {.};
        \path
            (cv1) edge[cvline] (cv5)
            (lll1) edge[doubleline] (ll2)
            (ll2) edge[doubleline] (l1)
            (ll1) edge[doubleline] (l2)
            (l2) edge[doubleline] (r1)
            (l1) edge[doubleline] (r2)
            (lll3) edge[] (ll3)
            (ll3) edge[] (l3)
            (l3) edge[] (r3)
            (r3) edge[] (rr3)
            (lll2) edge[] (ll2)
            (lll2) edge[doubleline] (ll1)
            (ll2) edge[] (l2)
            (l2) edge[] (r2)
            (lll2) edge[] (ll3)
            (lll3) edge[] (ll2)
            (ll2) edge[] (l3)
            (ll3) edge[] (l2)
            (l2) edge[] (r3)
            (l3) edge[] (r2)
            (r2) edge[] (rr3)
            (r2) edge[doubleline] (rr1)
            ;
\end{tikzpicture}
\end{equation}
Measuring the full input mode (the rightmost one) in the momentum basis involves projecting onto $\ketbra {s} {s} [p] [p]$, with outcome~$s$. For illustrative simplicity, we restrict to the outcome~$s=0$, but any outcome will have the same effect after correcting for the measurement-induced displacement~\cite{menicucci_universal_2006}. This measurement teleports the input state by one ``hop'' along the CV quantum wire and applies a Fourier transform, giving the following result:
\begin{equation}\label{tikz:fullmeasurement}
\begin{tikzpicture}[xscale=0.8,yscale=0.75]
        \node[cvnode] (cv1) at (-9,2) {};
        \node[cvnode] (cv2) at (-8,2) {};
        \node[cvnode] (cv3) at (-7,2) {};
        \node[emptycvnode] (cv4) at (-6,2) {};
        \node[] () at (-6,2) {{\footnotesize $\tilde \psi_{\GKPsmall}$}};
        \node[] () at (-4.1,2) {$=$};
        \node[rectanglenode] (lll1)     at (-3.3,1) {};
        \node[circlenode] (lll2)        at (-3.3,2) {};
        \node[logicalnode] (lll3)       at (-3.3,3) {};
        \node[rectanglenode] (ll1)      at (-2.3,1) {};
        \node[circlenode] (ll2)         at (-2.3,2) {};
        \node[logicalnode] (ll3)        at (-2.3,3) {};
        \node[rectanglenode] (l1)       at (-1.3,1) {};
        \node[circlenode] (l2)          at (-1.3,2) {};
        \node[logicalnode] (l3)         at (-1.3,3) {};
        \node[rectanglenode] (r1)       at (-0.3,1) {};
        \node[emptycirclenode] (r2)     at (-0.3,2) {};
        \node[emptylogicalnode] (r3)    at (-0.3,3) {};
        \node[] ()            at (-0.3,3)          {{\footnotesize $\tilde \psi$}};
        \node[] () at (1.2,1.9) {.};
        \path
            (cv1) edge[cvline] (cv2)
            edge[cvline] (cv3)
            edge[cvline] (cv4)
            (lll1) edge[doubleline] (ll2)
            (ll2) edge[doubleline] (l1)
            (ll1) edge[doubleline] (l2)
            (l2) edge[doubleline] (r1)
            (lll3) edge[] (ll3)
            (ll3) edge[] (l3)
            (l3) edge[] (r3)
            (lll2) edge[] (ll2)
            (lll2) edge[doubleline] (ll1)
            (ll2) edge[] (l2)
            (lll2) edge[] (ll3)
            (lll3) edge[] (ll2)
            (ll2) edge[] (l3)
            (ll3) edge[] (l2)
            (l2) edge[] (r3)
            ;
\end{tikzpicture}
\end{equation}
The physical state at the rightmost mode is $\ket*{\tilde \psi}[\GKP] = \op F \ket\psi[\GKP]$, and the logical state is now~$\ket*{\tilde\psi}[L] = \op H_L \ket \psi[L]$. This single step implements a Fourier transform~$\op F$ at the physical level and a Hadamard gate~$\op H_L$ at the logical level, which is exactly what would have happened if the original state were a completely GKP cluster state. Why did the initial gauge-mode entanglement [seen in \cref{tikz:CVQW}] not spoil this connection?

To interpret how this has happened, we note that $\ketbra {0}{0}[p][p] = \ketbra {+}{+}[L][L] \otimes \ketbra {0}{0}[p,G][p,G]$. This means that we can interpret the physical $p$~measurement as two separable measurements, one on each subsystem. The logical measurement is exactly what we would expect for this case: a measurement in the $\op X_L$ direction of the logical Bloch sphere with outcome~$+1$, and the gauge mode is measured in~$p_G$. Since these act on different subsystems, we can pretend that they had been done \emph{sequentially} even though they were actually simultaneous. (The value of this will become clear in a moment.)

We consider the gauge-mode measurement as happening first, followed by the logical one. In order to figure out what the intermediate graph (``between'' the two parts of the full measurement) would look like, we start with what we know. First, we know that the measured gauge mode must have disappeared, along with any of its adjacent edges. Second, we can work backwards: We know that whatever graph we draw here has to be the precursor to the graph in \cref{tikz:fullmeasurement}---it must produce that graph after a measurement of $\op X_L$ with outcome~$+1$. Furthermore, this has to be the case \emph{regardless of the input~$\ket \psi[L]$}. That constrains the result to be the following:
\begin{equation}
\begin{tikzpicture}[xscale=0.8,yscale=0.75]
        \node[cvnode] (cv1) at              (-9,2) {};
        \node[cvnode] (cv2) at              (-8,2) {};
        \node[cvnode] (cv3) at              (-7,2) {};
        \node[cvnode] (cv4) at              (-6,2) {};
        \node[mgkp] (cv5) at                (-5,2) {};
        \node[text=amaranth] () at          (-5,2)
             {{\footnotesize $\psi_{\GKPsmall}$}};
        \node[] () at                       (-4.1,2) {$=$};
        \node[rectanglenode] (lll1) at      (-3.3,1) {};
        \node[circlenode] (lll2) at         (-3.3,2) {};
        \node[logicalnode] (lll3) at        (-3.3,3) {};
        \node[rectanglenode] (ll1) at       (-2.3,1) {};
        \node[circlenode] (ll2) at          (-2.3,2) {};
        \node[logicalnode] (ll3) at         (-2.3,3) {};
        \node[rectanglenode] (l1) at        (-1.3,1) {};
        \node[circlenode] (l2) at           (-1.3,2) {};
        \node[logicalnode] (l3) at          (-1.3,3) {};
        \node[rectanglenode] (r1) at        (-0.3,1) {};
        \node[emptycirclenode] (r2) at      (-0.3 0,2) {};
        \node[logicalnode] (r3) at          (-0.3,3) {};
        \node[mrectangle] (rr1) at          (0.7,1) {};
        \node[mcircle] (rr2) at             (0.7,2) {};
        \node[emptylogicalnode] (rr3) at         (0.7,3) {};
        \node[] () at             (0.7,3)
            {{\footnotesize $\psi$}};
        \node[] () at ( 1.2,1.9) {,};
        \path
            (cv1) edge[cvline] (cv5)
            (lll1) edge[doubleline] (ll2)
            (ll2) edge[doubleline] (l1)
            (ll1) edge[doubleline] (l2)
            (l2) edge[doubleline] (r1)
            (lll3) edge[] (ll3)
            (ll3) edge[] (l3)
            (l3) edge[] (r3)
            (r3) edge[] (rr3)
            (lll2) edge[] (ll2)
            (lll2) edge[doubleline] (ll1)
            (ll2) edge[] (l2)
            (lll2) edge[] (ll3)
            (lll3) edge[] (ll2)
            (ll2) edge[] (l3)
            (ll3) edge[] (l2)
            (l2) edge[] (r3)
            ;
\end{tikzpicture}
\end{equation}
where the dotted-outlined node on the left represents that the node has had its gauge mode measured destructively (technically, an unphysical operation), with analogous notation on the right.
From this intermediate step, we finally have some intuition.

Notice that there is an open circle, representing $\ket{u_G = 0}[G]$, where there used to be a filled one [in \cref{tikz:CVQW}].  Also recall that that open circle will show up regardless of what $\ket\psi[L]$ is. Thus, it will also be there even when $\ket \psi[L] = \ket 0[L]$, which disconnects the logical input from the rest of the initial state in \cref{tikz:CVQW}. This means that the changing of the closed circle to an open one cannot be a result of entanglement with the logical input state since it must happen even if there is no such entanglement ($\ket \psi[L] = \ket 0[L]$). Thus, the only place it could have come from is the original (measured) gauge mode. We can narrow down its origin even further by noticing that the $u_G$ piece of the gauge mode was originally an open circle in \cref{tikz:CVQW}, which means it was disconnected from everything else and thus could not have had any effect on the output state through being measured.

So now we finally have the resolution to the original puzzle: \emph{Upon measurement of~$\op p$ of the physical mode, the filled square in the measured gauge mode gets teleported into the open circle of the next gauge mode.} The map between these two nodes is a type of generalized Fourier transform, examples of which are ubiquitous in cluster-state quantum computation~\cite{raussendorf_one-way_2001,menicucci_universal_2006,albert2020robust}. In this case, it is specifically a Fourier series that relates the two since a $\delta$-function in a compact interval (open circle) has a uniform Fourier series over all integers (filled square).
Interpreting these subsystems as two rotors (see Sec.~\ref{sec:ssd}), makes this clear, since a straightforward unitary Fourier relation exists between the angular momentum basis and the angular position basis~\cite{raynal_encoding_2012}.

This replacement of the closed circle by an open one eliminates all of the edges emanating from that node, thereby disconnecting (and disentangling) it entirely from the rest of the graph. Due to its visual similarity to a zipper separating the upper and lower layers of connecting teeth, we refer to this process informally as ``unzipping'' the cluster state. It is the specific form of the GKP state's gauge mode that makes this unzipping possible.

At the full-mode level, all that is happening is that a GKP state gets teleported to the next mode and undergoes an $\op F$ gate. But at the subsystem level, the behavior is much richer and more intricate. From this perspective, the filled square---which represents the repetitive nature of the GKP state in position space---``unzips'' the cluster state by disconnecting the measured logical qubit from all gauge modes so that the logical information can teleport one step as if there were no connections to any gauge modes from the beginning. This unzipping lets the information processing of GKP states in a CV or hybrid cluster state [Fig.~\ref{fig:graphicaldecompositions}(a,c)] mimic that happening in a GKP one [Fig.~\ref{fig:graphicaldecompositions}(b)] since at each step the gauge modes are disconnected before the logical information moves forward. The process of repeated measurement of the full modes can thus be pictured as repeatedly unzipping by one step, then doing ordinary logical-qubit teleportation, then repeating this pattern all the way down the CV quantum wire.

Of course, there is another way of changing a filled circle into an open one, and that is to do GKP error correction on the $q$ quadrature. In a way, this shows that the unzipping process involves using the gauge mode of the input to GKP error correct that of the next mode in one quadrature, thereby allowing the input logical information to teleport onward, untarnished by entanglement with its neighbors.

The purpose of going through this exercise is to illustrate the conceptual insights that the SSD can bring to questions that would otherwise appear mysterious. In this example, the question we answered is why CVCSs play so nicely with GKP qubits, a fact first discovered in Ref.~\cite{menicucci_fault-tolerant_2014}. Such intuition has recently informed further proposals for other hybrid cluster states, such as Ref.~\cite{bourassa2021blueprint}, which shares two authors with the present work.

\section{Finite squeezing}
\label{sec:fsqueez}

In the above analyses, we have focused on cluster states built from ideal 0-momentum states and ideal GKP states. Each of these is the asymptotic limit of a physically realizable state that contains some inherent imperfections in the form of finite squeezing.  Nevertheless, the ideal, infinitely squeezed states have allowed us to discuss the entanglement structure and the fundamentals of quantum information propagation for several types of cluster state.  From this perspective, it has been fruitful to analyze these idealized states, as they embody the essential features that make cluster states useful for quantum computation~\cite{zhang_continuous-variable_2006, zhang_graphical_2008}.

The physical counterparts to the ideal cluster states we have considered throughout this work are obtained by coupling modes with the same CV controlled-$Z$ entangling operator, \cref{eq:controlledzdecomposed}, but with the modes prepared in finitely squeezed approximations to a 0-momentum eigenstate~\cite{menicucci_universal_2006} or to an ideal GKP state~\cite{gottesman_encoding_2001, matsuura2020equivalence}. This finite squeezing makes the cluster states inherently noisy, and this noise compounds during quantum computation, necessitating error correction in order to achieve fault tolerance~\cite{menicucci_fault-tolerant_2014}.  The reader interested in the general implications of finite squeezing for CVCSs and GKP states can find an extensive investigation in Ref.~\cite{pantaleoni2021}.
In this section, we discuss the finite-squeezing generalization of our results from a graph-rule perspective.

The first generalization would be to include momentum-squeezed vacuum states, which approximate 0-momentum eigenstates to a degree determined by the level of squeezing. In previous work~\cite{pantaleoni2021}, we performed the modular-position SSD of a squeezed vacuum state by first writing it as $0$-momentum states distorted by a squeezing-dependent \emph{envelope} operator;
$
\ket{0,\zeta}[p] \propto \exp\pqty{ - \zeta^2 \q^2 /2 }\ket{0}[p]
$ for some real $\zeta$ describing the amount of squeezing~\cite{walshe_2020}. In the SSD, the envelope operator contains interaction terms that entangle the gauge and logical subsystems of the $0$-momentum state. These interactions arise from the exponentiation of $
\q^2 = (\alpha \l + 2 \alpha \mGop + \uGop)^2
$, which contains cross terms between the logical and gauge subsystems of the same physical mode---for example, terms proportional to $\l \otimes \uGop$, as well as self-interaction terms like $\uGop^2$.

While the graphical rules we introduce in \cref{sec:clusterstates} are not general enough to describe either of these types of interaction, we can take inspiration from the graphical calculus for Gaussian pure states~\cite{menicucci_graphical_2011}, which makes use of arbitrary-weight, complex-valued graphs and self loops to describe pure Gaussian states such as finitely squeezed CVCSs.  We may be able to import these innovations to describe the envelope operator in the SSD by (a)~generalizing edges connecting two modes to arbitrary-weight, complex-valued interactions, and (b)~allowing for complex-valued self-loops.

We have shown above that the SSD provides a surprisingly simple graphical description of several extremely non-Gaussian states: ideal GKP states and cluster states built from them. However, just as for 0-momentum states, physical realizations of GKP states contain inherent imperfections (in the form of finite squeezing of the spikes) that are not present in their idealizations.\footnote{One may argue that ideal GKP states are \emph{more} unphysical than 0-momentum states, as they are defined as an infinite sum of unphysical wave functions.}
Thus, a second important generalization to our graphical SSD description would be to include finitely squeezed GKP states, also commonly referred to as \emph{approximate GKP states}.
However, unlike squeezed vacuum states discussed above, approximate GKP states cannot be written simply as a position-space envelope operator acting on an ideal GKP state. In fact, there are a number of different ways to approximate GKP states~\cite{matsuura2020equivalence}, none of which admits an obvious graphical representation, even at the CV-mode level (before the SSD is performed). An SSD-based analysis of one common parametrization of approximate GKP states can be found in Ref~\cite{pantaleoni2021}.

A common approach to approximate high-quality---
\emph{i.e.},~high-squeezing---GKP states replaces each periodically placed $\delta$ function in an ideal GKP wave function (both position and momentum) with a Gaussian spike whose width is determined by the level of squeezing. Additionally, a broad Gaussian envelope with variance proportional to the inverse of the squeezed variance damps out Gaussian spikes far from the origin.  One possible way forward towards a graphical representation of these states at the CV-mode level could be to leverage the fact that the wave functions are characterized by a single covariance matrix~\cite{matsuura2020equivalence, mensen2020phasespace}. This could potentially be combined with another feature of the graphical calculus for Gaussian states: the explicit connection between a covariance matrix and graphical representation of the state.
Going beyond a CV-mode to an SSD description, both the spike width and the broad envelope produce nontrivial entangling effects between the virtual subsystems, again creating challenges to a simple graphical description.  We expect that additional graphical innovations will be required in order to account for the inherent noise in approximate GKP states.

When applying the new graphical formalism to practical problems, even more innovation will be required. As an example, consider the evolution of GKP-encoded information as it teleports along a CV quantum wire. This process would also need to be modified to include more general graph transformation rules, analogous to the generalization of Zhang's rules for transforming CVCS graphs among themselves~\cite{zhang_local_2008,zhang_graphical_2008} to those for transforming any Gaussian pure state under Gaussian unitaries~\cite{menicucci_graphical_2011}. The tools developed along the way will likely be of use for quantitative analysis and comparison of performance of the different types of cluster state (\cref{fig:graphicaldecompositions}). Since it represents a significant extension beyond the present paper, however, we leave this generalization to future work.

\section{Conclusion}

Expanding on the graphical formalism already introduced in~\cite{pantaleoni_modular_2020}, we have shown that a hybrid cluster state, such as the ones considered in the recent proposal for continuous-variable quantum computing~\cite{bourassa2021blueprint}, presents an entanglement structure that mimics the behavior of a discrete-variable cluster state (upon measurements). The subsystem decomposition is the mathematical formalism that has allowed us to draw this connection in an explicit way. Here we summarize the main points of the investigation.

As a starting point, we have decomposed the fundamental entangling gate, a two-mode CV controlled-$Z$ operator.  We found that it contains a qubit controlled-$Z$ operation on the logical subsystems of the two modes as well as additional pieces that entangle the qubit subsystems to the gauge modes. When properly tuned, some of these unwanted interactions vanish.

From the subsystem perspective, the logical state of an ideal CVCS is exactly a discrete-variable cluster state that is entangled  with  the gauge subsystems of the rest of the modes. The interaction term in the decomposed controlled-$Z$ suggests that further steps need to be taken if we want the cluster state to implement exactly a logical-qubit cluster state that is separable from the gauge modes. GKP states turn out to be the additional ingredient that disentangles the logical cluster state from the gauge mode. This allows the resource state to behave effectively as a qubit cluster state, free from entanglement with the gauge mode, as discussed in \cref{sec:unzipping}. An ideal GKP state, when coupled to a CVCS, will ``unzip'' the hidden qubit cluster state from the gauge modes before allowing the logical information to teleport to the next node, thereby avoiding logical decoherence that would otherwise arise due to the initial entanglement.

The graphical SSD representation of ideal hybrid cluster states can be rigorously extended to finitely squeezed versions of these states. Such a development would mirror that of the graphical calculus for Gaussian pure states~\cite{menicucci_graphical_2011} that extended previously known graphical representations, which were limited to ideal CVCSs~\cite{gu_quantum_2009,zhang_graphical_2008}, and it would allow for quantitative modeling of the evolution of such states and of their use in processing CV quantum information.

\begin{acknowledgments}
This work was supported by the Australian Research Council Centre of Excellence for Quantum Computation and Communication Technology (Project No.~CE170100012).
\end{acknowledgments}

\vfill

\bibliography{references}
\bibliographystyle{apsrev4-2_title}

\end{document}